# Swansong Biospheres II: The final signs of life on terrestrial planets near the end of their habitable lifetimes

Jack T. O'Malley-James
School of Physics and Astronomy, University of St Andrews, North Haugh, St Andrews, Fife, UK.

Charles S. Cockell
UK Centre for Astrobiology, School of Physics and Astronomy, James Clerk Maxwell Building, The King's Buildings, University of Edinburgh, Edinburgh, UK.

Jane S. Greaves
School of Physics and Astronomy, University of St Andrews, North Haugh, St Andrews, Fife, UK.

John A. Raven
Division of Plant Sciences, University of Dundee at TJHI, The James Hutton Institute, Invergowrie, Dundee, UK.

**Corresponding author:**
J.T. O'Malley-James
School of Physics & Astronomy
University of St Andrews
North Haugh
St Andrews
Fife, KY16 9SS
E.mail: jto5@st-andrews.ac.uk

**Abstract**

*The biosignatures of life on Earth do not remain static, but change considerably over the planet's habitable lifetime. Earth's future biosphere, much like that of the early Earth, will consist of predominantly unicellular microorganisms due to the increased hostility of environmental conditions caused by the Sun as it enters the late stage of its main sequence evolution. Building on previous work, the productivity of the biosphere is evaluated during different stages of biosphere decline between 1 Gyr and 2.8 Gyr from present. A simple atmosphere-biosphere interaction model is used to estimate the atmospheric biomarker gas abundances at each stage and to assess the likelihood of remotely detecting the presence of life in low-productivity, microbial biospheres, putting an upper limit on the lifetime of Earth's remotely detectable biosignatures. Other potential biosignatures such as leaf reflectance and cloud cover are discussed.*

**INTRODUCTION**

The later stages of the Sun's main sequence evolution (2-3 Gyr from the present) will result in much higher surface temperatures on the future Earth and therefore, much more extreme environments for the last life able to grow and survive on the planet. This work builds upon earlier work (O'Malley-James *et al.* 2013) in which potential refuge environments for life on the future Earth were identified, in order to evaluate what remotely detectable biosignatures the Earth would produce as the Sun's main sequence evolution moves the inner edge of the habitable zone (HZ) outwards driving extreme climate change. The habitable lifetime of any planet is limited, the exact duration depending on the type of star (or stars) hosting that planet; hence, knowing the likely biosignature evolution of the dying Earth can inform us about the potential remote signature appearances of Earth-like exoplanets whose biospheres are dying. A habitable planet can be considered to become uninhabitable when it crosses the inner edge of the HZ, which is defined by the runaway greenhouse limit - the point at which the stellar flux reaching the top of a planet's atmosphere crosses a threshold value that triggers runaway heating. The most recent estimate places the inner edge of the Sun's HZ at 0.99 au (Kopparapu *et al.* 2013), corresponding to a threshold flux of $1.06\ S_0$ (where $S_0$ is the present solar flux). However, other work places higher limits on this threshold of up to $1.25\ S_0$ (Goldblatt & Watson 2012; Goldblatt *et al.* 2013).

The death of the biosphere as we know it today begins with the extinction of higher plant species. Rising temperatures cause silicate weathering rates to increase, increasing $CO_2$ draw-down, lowering $CO_2$ levels in the atmosphere. This results in conditions that are increasingly unsuited to (higher) plant life (Lovelock & Whitfield 1982; Caldeira & Kasting 1992). During the $CO_2$ decline, rapid ocean evaporation would not yet have begun. From Henry's Law, a reduction in atmospheric $CO_2$ would lead to a reduction in the $CO_2$ levels in the surface ocean, while increased silicate weathering could potentially lead to increased carbonate deposition. A reduced $CO_2$ concentration would lead to an increase in ocean pH, although organisms should be able to tolerate more alkaline oceans. Cyanobacteria have developed $CO_2$-concentrating mechanisms in response to the low affinity for $CO_2$, and the low selectivity of Rubisco (ribulose-1,5-bisphosphate carboxylase-oxygenase), the enzyme involved in the first step of the carbon fixation process) for $CO_2$ over $O_2$, which involves importing/accumulating inorganic carbon in the form of bicarbonates in the cytoplasm and converting this to $CO_2$ (Koropatkin *et al.* 2007; Zarzycki *et al.* 2013). Those that live in environments with variable inorganic carbon (*Synechocystis* PCC 6803, for example) have high-affinity bicarbonate transporters that are used under low $CO_2$ conditions using energy, ultimately from light, to accumulate $CO_2$ inside the cells to values much higher than those in the external medium (Koropatkin *et al.* 2007). Such mechanisms, although with lower concentration factors (inside concentration/outside concentration) for $CO_2$, occur in many eukaryotic algae (Raven *et al.* 2012). Mechanisms like this could allow oceanic photosynthesis to survive for longer than on land (Raven *et al.* 2012).

The continual decrease in $CO_2$ levels eventually renders photosynthesis impossible for higher plant species, bringing an end to the age of plants. A corresponding decrease in atmospheric oxygen levels, coupled with the loss of primary food sources, would lead to the concurrent, sequential extinction of animal species, from large vertebrates to smaller ones, with invertebrates having the longest stay of execution (O'Malley-James *et al.* 2013).

Life on Earth will then once again become microbial. Initially, while there is still sufficient atmospheric oxygen and carbon dioxide to fuel microbial metabolisms, this global microbial biosphere would likely be diverse, with maximum productivity values similar to those of the pre-photosynthetic Earth; $180-560 \times 10^{12}$ mol C yr$^{-1}$ (Canfield 2005). The actual productivity will probably be lower as primary productivity in the absence of photosynthesis involves chemolithotrophy, which would be subject to similar constraints on the supply of inorganic carbon as photosynthesis. When atmospheric oxygen reaches negligible levels, a rapid shift (within a few Myr) towards anaerobic metabolisers will occur. The possible microbial functional types that could grow under these conditions are discussed in [Table 1](). Chemolithotrophs, although they use a wider range of carboxylases than photosynthetic organisms, would be subject to similar carbon constraints to cyanobacteria because none of the carboxylases used have much higher affinity for inorganic carbon than the highest-affinity members of the family of photosynthetic carboxylases Rubisco (Raven *et al.* 2012).

Earth's surface becomes largely uninhabitable between 1.2 - 1.85 Gyr from present, depending on latitude (O'Malley-James *et al.* 2013), consistent with recent estimates of 1.75 Gyr for Earth's habitable lifetime by Rushby *et al.* (2013). However, as surface temperatures across the planet continue to increase, lower temperature refuges where liquid water is still available will still exist beyond this time. In O'Malley-James *et al.* (2013) potential refuge environments in the subsurface, caves and at high altitudes were discussed, which could enable a biosphere to exist for up to 2.8 Gyr from present. This biosphere would favour unicellular, anaerobic organisms with a tolerance for one or more extreme conditions.

In this paper, changes in biosignatures as the biosphere declines, from the extinction of animals and plants to an increasingly less diverse microbial world, are modelled and their potential for remote detection discussed.

**METHODS**

A simple atmosphere-biosphere chemical interaction model was used, combining the metabolic reactions listed in Table 1, abiotic gas fluxes and simple atmospheric chemistry. The model is linked to a surface temperature evolution model (shown in Figure 1), described in O'Malley-James *et al.* (2013), where the surface temperature

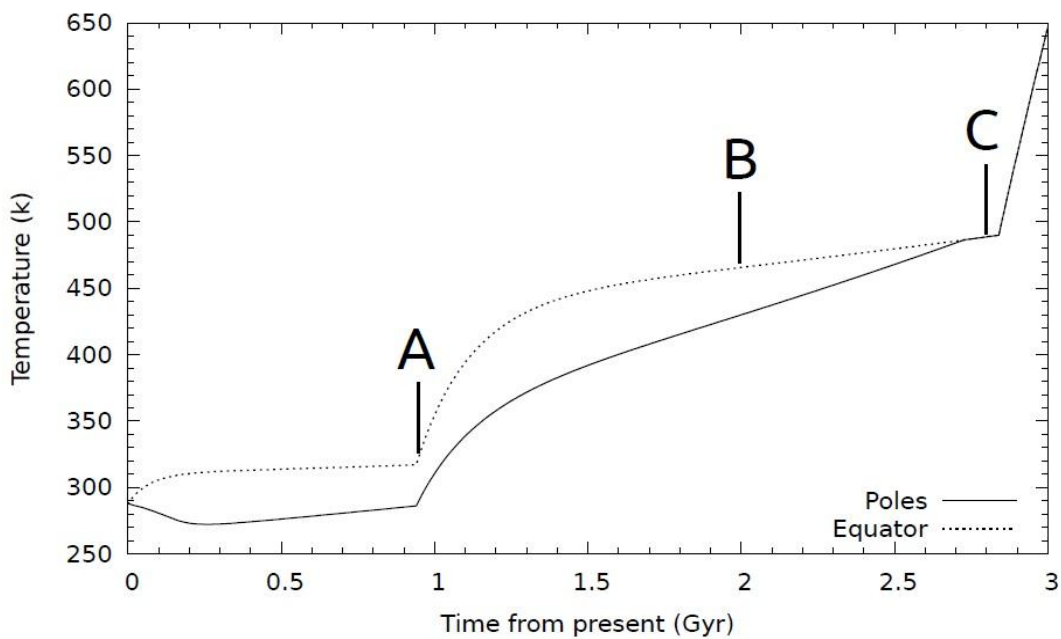

**Figure 1**: Points during the Earth's future temperature evolution at which biosignature gas levels are measured. Point A represents the period during which all plant and animal species become extinct; B represents a low-productivity microbial biosphere and C represents a biosphere of final survivors living in refuge environments before all life on Earth becomes extinct.

(T) evolution with latitude and altitude is modelled using

$$\frac{dT}{dt} = \frac{F_{in} - F_{out}}{C_p}$$

$F_{in}$ and $F_{out}$ represent the incoming and outgoing energy fluxes to/from the planet and $C_p$ is the heat capacity of the planet (at constant pressure). The model was run for 3 Gyr accounting for increased solar luminosity (L) using,

$$L(t) = \left[1 + \frac{2}{5}\left(1 - \frac{t}{t_e}\right)\right]^{-1} L_s$$

where t is the time elapsed on the main sequence, $t_e$ is the current age of the Sun and $L_s$ is the present-day solar luminosity. Long-term variations of eccentricity and obliquity are accounted for over the modelled time period as described in O'Malley-James *et al.* (2013). The modelled time period begins during the period in which all plants and animals become extinct (approximately 1 Gyr from the present; point A in Figure 1). A global microbial biosphere is assumed thereafter until those organisms can no longer survive.

This biosphere is assumed to consist of microbial species using the metabolic processes listed in Table 1. An initial microbial abundance similar to the total of present estimates for the terrestrial subsurface (in this case defined as habitats below 8 m in depth) ($2.5 \times 10^{29}$ cells), soils ($2.6 \times 10^{29}$) and sub-seafloor sediment ($2.9 \times 10^{29}$) (Whitman *et al.* 1998; Kallmeyer *et al.* 2012) was assumed. This microbial biosphere is initially evenly distributed over the surface of the Earth, giving an initial cell count of $1.5 \times 10^{15}$ cells m$^{-2}$. Splitting the planet into three latitude zones (polar (±60-90°), mid-latitude (±30-60°) and equatorial (0-±30°)) and assuming ~1% of the surface area is at an elevation ≥ 5 km gives the microbial abundance within the latitude zones assumed in the temperature model.

The microbial abundance in a particular region is then linked to the surface temperature evolution such that the count drops to zero when the temperature in a particular zone exceeds 420 K; an upper temperature bound for life (set to allow some increase over the currently known upper temperature tolerance of thermophiles) at which biological chemical processes break down). The total modelled cell abundance is shown in Figure 2.

The chemical reactions occurring in a single cell for each type of microbial metabolism (see Table 1) are modelled by calculating the per cell rate of the gas production/consumption (see Table 2) for species representing the different microbial metabolic types using data from studies in recent literature. By combining this information with the estimated microbial cell abundance, reactants are drawn from an atmospheric reservoir of gases and added to that reservoir, assuming the reactant use rates/product production rates described in Table 2. These rates are used to determine the production rates/reactant use rates in the metabolic reactions using the individual molar ratios of reactants to products in a given reaction. The limiting reactant (the one not in excess) determines the yield of products in each case.

| Microbe type | Likely Habitat | Reactions |
|---|---|---|
| Methanogens | High elevation, caves, subsurface. Most abundant before photosynthesising plants die off, but increasingly limited to $CO_2$ sites (e.g. volcanic vents) as atmospheric carbon dioxide levels drop. | $CO_2 + 4H_2 \rightarrow CH_4 + 2H_2O$<br>$CH_3COOH \rightarrow CH_4 + CO_2$<br>Other C sources, e.g. methanol<br>(Ferry 2002) |
| Ammonia oxidisers | High elevation, caves. Limited to volcanic vents; only until atmospheric oxygen drops to negligible levels. | $NH_3 + O_2 + 2e^- \rightarrow NO_2^- + 5H^+ + 4e^-$<br>(Bernhard 2012) |
| Hydrogen oxidisers | Subsurface. Anywhere where hydrogen is available. Photolysis of water vapour could provide abundant atmospheric hydrogen. | $2H_2 + O_2 \rightarrow 2H_2O$<br>$2H_2 + CO_2 \rightarrow (CH_2O) + H_2O$<br>(Bongers 1970) |
| Sulphur oxidisers | Caves. Limited to volcanic vents and evaporites. Only until atmospheric oxygen drops to negligible levels. May be able to persist near localised $O_2$ sources, e.g. near anaerobic iron reducers. | $2S + 3O_2 + 2H_2O \rightarrow 2H_2SO_4$<br>$12FeSO_4 + 3O_2 + 6HOH \rightarrow 4Fe_2(SO_4)_3$<br>(Suzuki $et~al.$, 1990; Suzuki $et~al.$, 1992) |
| Carboxydotrophs | High elevation, caves. Limited to near volcanic vents. Only until atmospheric oxygen drops to negligible levels. May be able to persist near localised $O_2$ sources. A reduction in $O_2$ levels favours the photodissociation of $CO_2$ to CO in the troposphere (Pinto $et~al.$, 1980), perhaps expanding the habitats for CO using microorganisms, however, this is also dependent on $CO_2$ availability. | $CO + O_2 + 2e^- + 2H^+ \rightarrow CO_2 + H_2O$<br>(Bender & Conrad 1994) |
| DMS oxidisers | High elevation, caves. Initially very abundant during the end of the age of photosynthesis when photosynthetic microorganisms still exist. Declining as photosynthetic life ends and oxygen levels fall. | DMS oxidation, producing: SO, dimethyl sulphoxide, dimethylsulphone, methane sulfonic acid, sulphuric acid<br>(Nicholas Hewitt & Davison, 1997) |
| Iron oxidisers | Abundant in all areas with liquid water and a source of iron until atmospheric oxygen levels fall to negligible levels. | $2Fe(OH)_2 + O_2 \rightarrow H_2O + Fe_2O_3$<br>$10FeCO_3 + 2NO_3^- + 24H_2O \rightarrow 10Fe(OH)_3 + N_2 + 10HCO_3^- + 8H^+$<br>(Straub $et~al.$ 1996) |
| Aerobic Methanotrophs | Caves, subsurface. Most abundant immediately after photosynthesising plants die off, but increasingly limited to $CH_4$ sites (e.g. volcanic vents) as atmospheric $CH_4$ levels drop. | $CH_4 + 2O_2 \rightarrow CO_2 + 2H_2O$<br>(Mancinelli 1995) |
| Anaerobic Methanotrophs | Subsurface, caves, high elevation. Most abundant immediately after photosynthesising plants die off, but increasingly limited to $CH_4$ sites (e.g. volcanic vents) as atmospheric $CH_4$ levels drop. | $CH_4 + SO_4^{2-} + 2H^+ \rightarrow CO_2 + 4H_2 + H_2S$<br>(Mancinelli 1995) |
| Iron reducers | Subsurface, caves. Abundant in any place with liquid water and a source of iron. Likely to be able to survive until temperatures exceed the maximum temperature tolerance of life. | Carbohydrate $e^-$ donor: $2Fe_2O_3 + 3CH_2O \rightarrow 4Fe + 3CO_2 + 3H_2O$<br>Hydrogen $e^-$ donor: $2Fe_2O_3 + 3H_2 \rightarrow 2Fe + 3H_2O$<br>Aquifer soils: $H_2O + Fe_2O_3 \rightarrow 2Fe(OH)_2 + O_2$ |

| | | (Sawyer et al. 1967; Longbottom & Kolbeinsen 2008) |
|---|---|---|
| Sulphate reducers | Subsurface, caves, high elevation. Limited to volcanic vents and evaporites. Likely to be able to survive until temperatures exceed the maximum temperature tolerance of life. | $CH_4 + SO_4^- + 2H^+ \rightarrow H_2S + CO_2 + 2H_2O$ (Barnes & Goldberg 1976) |
| Acetogens | Subsurface, caves. Most abundant during before plants die off. Limited to $CO_2$ and $H_2$ sources thereafter, perhaps associated with other microorganisms (e.g. $H_2$ or $CO_2$) producers. | $CO_2$ reduced to CO, converted to acetyl coenzyme A |
| Anammox Reaction | Subsurface, caves, high elevation. | $NH_4^+ + NO_2^- \rightarrow N_2 + 2H_2O$ (Strous et al. 2006) |
| Denitrification | Subsurface, caves | $2CH_2O + 2NO_3^- + 2H^+ \rightarrow 2CO_2 + N_2O + 3H_2O$ $5CH_2O + 4NO_3^- + 4H^+ \rightarrow 5CO_2 + 2N_2 + 7H_2O$ |
| Ammonia fermentation | Subsurface, caves | $2C_2H_5OH + NO_3^- + H_2O \rightarrow 2CH_3COOH + NH_4^+ + 2OH^-$ (Zhou et al. 2002) |

**Table 1:** Examples of microbial organisms that could be supported by environments on the far-future Earth. If organisms are to benefit (i.e. increase biomass), reactions have to be coupled to inorganic carbon uptake and reduction. Increasingly, conditions will favour anaerobic extremophiles until temperatures rise beyond the maximum temperature tolerance of life. The first seven rows, and the last row, represent organisms which are chemolithotrophs; the other rows represent chemoorganotrophs (= heterotrophs). [N.B. Ammonia oxidisers typically produce nitrite, and nitrite oxidisers produce nitrate; both are termed nitrifiers. Nitric oxide is a minor by-product in ammonia oxidation].

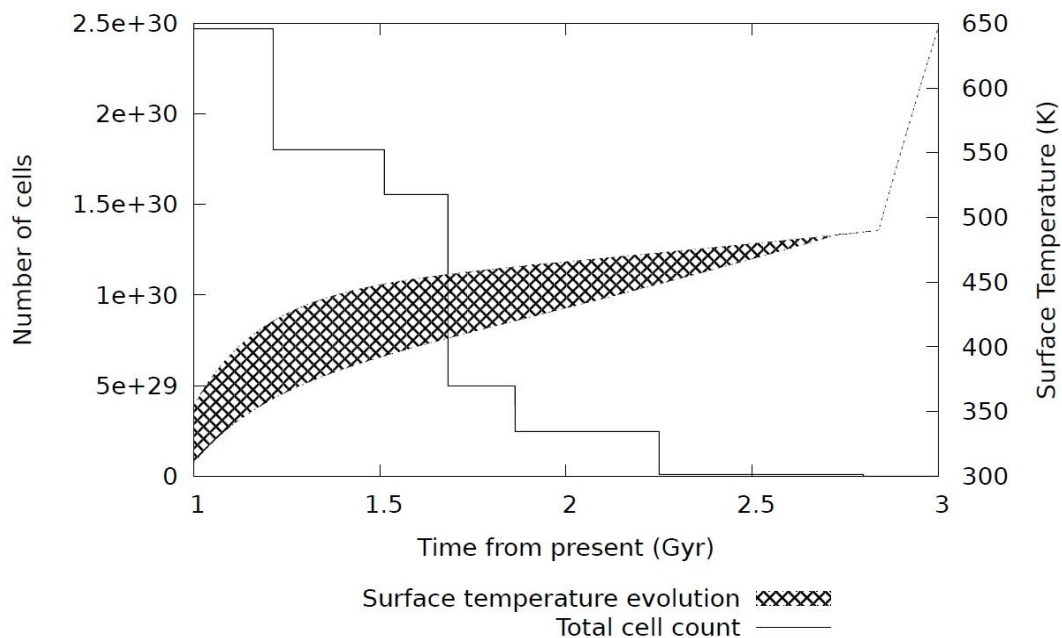

**Figure 2**: Change in cell abundance per unit surface area over time with increasing temperature. After ~2.2 Gyr, cell counts fall to ~$10^{28}$ cells until all life ends 2.8 Gyr from present. The shaded area represents the equator-to-pole temperature range.

The number of reactions taking place is determined by the temperature-dependent cell abundance calculated previously. The number of organisms using particular metabolisms is estimated based on the likely abundance of those metabolisers in the far-future environment. This is calculated from the total initial cell abundance estimate ($8 \times 10^{29}$ cells), assuming 1% are found at high altitude (> 5 km) and an order of magnitude fewer aerobic metabolisers than anaerobic metabolisers due to the low oxygen levels on the far-future Earth (based on abundance differences between aerobes and anaerobes in low oxygen environments, e.g. Tartakovsky *et al.* 2005; Ulloa *et al.* 2012), giving initial total cell counts of $8 \times 10^{28}$ cells and $7.2 \times 10^{29}$ cells for aerobes and anaerobes respectively. A "per-habitat" cell count for anaerobes and aerobes is calculated from these values (assuming 1% at high altitude and an even split between the cave and subsurface environments for the remainder in each case) and the "likely habitat" estimates from Table 1 are used to calculate an initial cell abundance for each of the modelled metabolisms (see Table 2). Other factors accounted for in the model (see Figure 3) are abiotic gas fluxes (see Table 3), ocean evaporation rates (1.32x10$^{15}$ g yr$^{-1}$ after the onset of rapid ocean evaporation) and hydrogen escape into space. While hydrogen escape on Earth is presently limited by diffusion through the homopause, during the rapid ocean evaporation phase the temperature-pressure structure of the atmosphere is altered, allowing large amounts of water vapour to reach the stratosphere where it photo-dissociates, resulting in a hydrogen-rich atmosphere. During this phase, hydrogen escape is limited only by the solar extreme UV heating rate (which, for an ageing sun-like star, would be similar in intensity to the present-day flux from the Sun and would not vary significantly over the modelled time period - Ribas *et al.* 2005); hence, a hydrodynamic escape rate of 3.06x10$^{30}$ H atoms s$^{-1}$ (Calderia & Kasting 1992) is assumed for the hydrogen-rich phase. At other times a diffusion-limited rate is assumed.

| Microbe type | Rate of gas consumption/production | | Source | Initial modelled cell count |
|---|---|---|---|---|
| | **Rate from source** | **Rate (g cell$^{-1}$ yr$^{-1}$)** | | |
| Ammonia oxidisers | *Nitrosomonas* species: 28.1 pmol - 0.2 fmol NH$_3$ oxidised cell$^{-1}$d$^{-1}$ Produces: NO | 3x10$^{-11}$ - 4x10$^{-6}$ | Boyd *et al.* (2011) | 8.0x10$^{27}$ |
| Carboxydotrophs | *Roseobacter* species: Max. of 1.1-2.3x10$^{-10}$ nmol CO oxidised cell$^{-1}$h$^{-1}$ | (2.7-5.6)x10$^{-14}$ | Tolli *et al.* (2006) | 8.0x10$^{27}$ |
| Methanogens | Wetland study: 273-665 µg CH$_4$ produced per kg soil per day with 1.07-8.29x10$^9$ cells per gram soil | (2.9-9.1)x10$^{-14}$ | Liu *et al.* (2011) | 9.9x10$^{28}$ |
| Sulphur oxidisers | *Thiooxidans* species: (2.5-9.9)x10$^{-4}$ µg S oxidised cm$^{-2}$ d$^{-1}$ | (0.9-3.6)x10$^{-18}$ | Smith *et al.* (2012) | 6.0x10$^{27}$ |
| Anaerobic methanotrophs | Reaction chamber: 8-33 nmol CH$_4$ consumed per gram sediment d$^{-1}$ | (4.7-19.4)x10$^{-14}$ | Girguis *et al.* (2003) | 9.9x10$^{28}$ |
| Aerobic methanotrophs | Landfill sites: 3-6.4 mmol CH$_4$ consumed per kg soil d$^{-1}$ | (1.8-23)x10$^{-11}$ | Kallistova *et al.* (2005) | 2.4x10$^{28}$ |
| Anaerobic iron reducers | Anaerobic sediment sites: 9-130 nmol Fe reduced per g sediment h$^{-1}$ Produces: CO$_2$ (Other iron reduction products: H$_2$O or O$_2$ (aquifer soils; Sawyer *et al.*, 1967), depending on electron donor; NB: using H$_2$O as an electron donor required energy input). | (4.4-64)x10$^{-12}$ | Sørensen (1982) | 8.1x10$^{28}$ |
| Aerobic iron oxidisers | *Leptospirillum ferrooxidans*: 10$^{-5}$ µmol Fe oxidised cell$^{-1}$d$^{-1}$ No gases produced | 2.1x10$^{-7}$ | Schrenk *et al.* (1997) | 2.6x10$^{28}$ |
| Hydrogen oxidisers | Isolates of soil hydrogen oxidising bacteria: 0.08-0.92 µmol H$_2$ oxidised h$^{-1}$ cm$^{-3}$ | (1.4-20)x10$^{-12}$ | Maimaiti *et al.* (2007) | 1.8x10$^{28}$ |
| Anammox | Anoxic water column site: (1534-2228)x10$^{-9}$ mmol N cell$^{-1}$ yr$^{-1}$ | (2.0-3.0)x10$^{-11}$ | Dalsgaard *et al.* (2003) | 9.9x10$^{28}$ |

**Table 2:** Consumption/production rates of biosignature gases from model species representing favourable metabolic pathways on the far-future Earth with calculated per-cell consumption/production rates and initial abundance estimates. If production rates are unknown, they are estimated based on consumption rates.

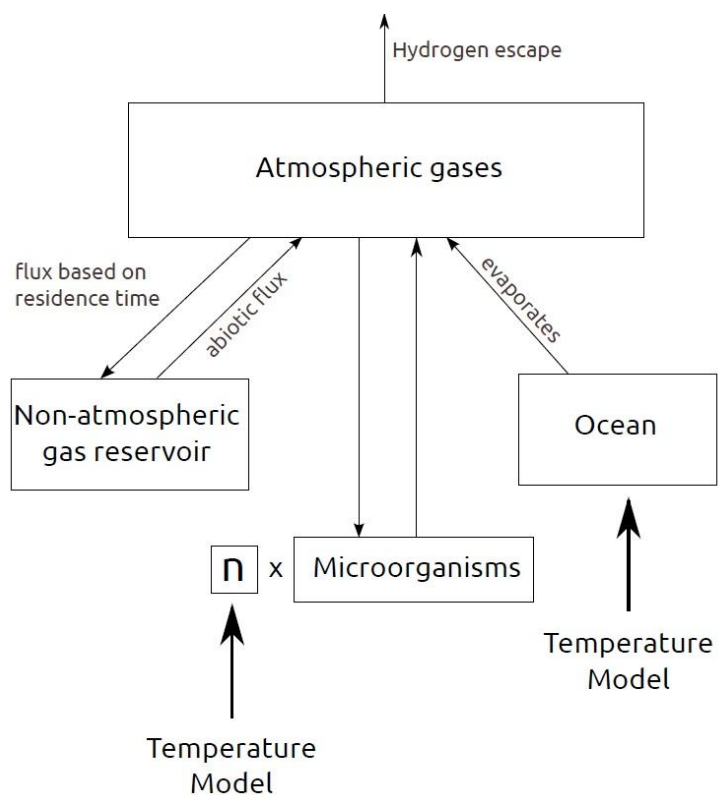

**Figure 3**: Schematic of the biosphere-atmosphere gas exchange model. The temperature model determines the abundance of organisms, n and the rates of gas fluxes between the atmosphere, ocean and a non-atmospheric gas reservoir consisting of the surface and subsurface. The abundance of specific types of microorganism is weighted based on the favourability of particular metabolisms on the far-future Earth.

The gas reservoir is then sampled at different time points (A, B and C in [Figure 1](#)) and biosignature gases are compared to their present-day abundances and their detectability evaluated.

| Gas | Present-day Flux to atmosphere | | Source |
|---|---|---|---|
| | **Flux (g yr$^{-1}$)** | **Gas Source** | |
| $CH_4$ | $5.3 \times 10^{13}$ | Volcanoes<br>Mud volcanoes<br>Hydrocarbon seepage | Burton *et al.* (2013) |
| $CO_2$ | $3 \times 10^{14}$<br>$3 \times 10^{14}$ | Volcanoes<br>Degassing | Burton *et al.* (2013) |
| $NO_2$ | $7 \times 10^{12}$ | Fixation by lightning | Tie *et al.* (2002) |
| $SO_2$ | $(15-21) \times 10^{12}$ | Volcanoes | Halmer *et al.* (2002) |
| $H_2O$ | $6.5 \times 10^{14}$ | Volcanoes | Fischer (2008) |
| CO | $CO:CO_2$ flux of 0.03-0.12<br>$2-3 \times 10^{14}$ | Volcanoes<br>Photochemical production | Wardell *et al.* (2004)<br>Zuo & Jones (1996) |

**Table 3**: Modelled abiotic gas fluxes. Given reduced tectonic activity on the far-future Earth, volcanic gas fluxes are reduced to ~20% of the present-day fluxes to account for the assumption that the only remaining volcanoes are hotspot volcanoes, which make up ~20% of present-day active subaerial volcanoes. The photochemical CO flux is linked to the dissolved organic carbon content of atmospheric water droplets and so is assumed to be negligible once plant and animal life becomes extinct.

**RESULTS & DISCUSSION**

*(i) Atmospheric biosignatures caused by the end of photosynthesis and the extinction of animals*

*a) Plants*

The end of the higher plants will probably come at the end of a gradual decline in plant diversity as species become less able to survive under changing environmental conditions. The net primary productivity (essentially the difference between photosynthesis and plant respiration) is 104.9 Pg C yr$^{-1}$, the terrestrial component of which is 56.4 Pg C yr$^{-1}$ (Field *et al.* 1998). At present 500 Pg of carbon is locked up in terrestrial plant biomass (Mahli 2002); hence, at present rates of primary productivity, this biomass could be produced in < 10 years. However, other loss processes result in the net increase in plant biomass being ~1.4 Pg C yr$^{-1}$ (Mahli 2002), giving an accumulation time of < 500 years at present rates.

Decaying plant matter initially transfers carbon to the soil carbon pool (1500 Pg C; Mahli 2002), microorganisms and, in some cases, to animals. From here it is mostly released to the atmosphere via decomposition or metabolism ($CO_2$ aerobically and $CH_4$ anaerobically), with other gases produced as a result of converting some of the energy in plant biomass into a form usable in cell synthesis and maintenance, increasing decomposer biomass (Allison 2006). Some carbon is transferred to oceans and geological carbon pools. Assuming similar carbon storage in the far-future biosphere, the death of all plants could release up to 500 Pg of carbon into the atmosphere via decomposition or animal metabolism over the time taken for plant species to decline. Higher temperatures are known to increase the rate of decay of dead plant matter into the soil carbon pool, with higher temperatures increasing the rate at which organic matter is broken down (Mahli 2002).

At present, 80% of the methane flux to the atmosphere comes from the decay of organic matter, estimated as $529\text{-}825 \times 10^{12}$ g $CH_4$ yr$^{-1}$ (Ehhalt 1974). The total dead phytomass (not including soil organic carbon) from which this is produced is approximately $184 \times 10^{15}$ g; approximately 15% of the estimated global total living phytomass, $1243.9 \times 10^{15}$ g (Ajtay *et al.* 1979). In the extreme case, in which the decline in plant life is very rapid, there would be an order of magnitude increase in methane flux until all dead phytomass has decayed. As methane has a relatively short atmospheric lifetime of approximately ten years (Kaltenegger *et al.* 2007) due to reactions with hydroxyl radicals and $O(^1D)$, there would be a more intense methane signature in Earth's atmospheric spectrum associated with this event (see [Figure 4](#)), but only for a short time period. This period may be shortened by increased production of hydroxyl radicals as a result of the increased atmospheric water vapour content.

However, the plant extinction sequence is likely to be less abrupt. For example, larger species tend to be less tolerant to increased temperatures compared to smaller species. The $CO_2$ uptake also varies for different plant types (Mooney 1972), suggesting that those with higher $CO_2$ uptake rates would be more vulnerable to extinction in a declining $CO_2$-environment.

Plants using the $C_3$ pathway to fix carbon (the dominant pathway in higher plants) would be able to survive until atmospheric $CO_2$ levels drop to 150 p.p.m.; 0.5 Gyr from now (Lovelock & Whitfield 1982; Caldeira & Kasting 1992; Raven *et al.* 2012).

Plant species may be able to continue to exist by becoming able to use lower concentrations of $CO_2$ for photosynthesis (Raven *et al.* 2012) and to tolerate arid environments or by implementing other nutritional strategies such as carnivory (Givnish *et al.* 1984; Rice 2002; Porembski *et al.* 2006; Raven *et al.* 2009; Albert *et al.* 2010; Flynn *et al.* 2013; Schmidt *et al.* 2013) or mycoheterotrophy (Leake & Cameron 2010; Schmidt *et al.* 2013). As a survival strategy, carnivory may be useful in certain low-nutrient ecosystems with intense solar radiation where liquid surface water is still abundant (Givnish *et al.* 1984; Rice 2002). These conditions would be more common near the beginning of the moist greenhouse stage as the planet begins heating up, e.g. loss of soil nitrogen in wet, high temperature conditions (although carnivory may be generally more related to phosphorus deficiency: Wakefield *et al.* 2005). However, $CO_2$ is still fixed from the atmosphere or freshwater bodies in carnivorous plants today, with the smallest fraction of carbon obtained from photosynthesis rather than carnivory in terrestrial *Utricularia* species (Porembscki *et al.* 2006; Albert *et al.* 2010) and almost all use the $C_3$ pathway to fix carbon, which does not permit rapid photosynthesis in very low $CO_2$ concentration, high-temperature environments. Some bromeliads (a few of which may be carnivorous: Givnish *et al.* 1984) use CAM to fix carbon, which helps to acquire and conserve carbon in arid environments. However, ultimately, as a strategy, carnivory would only give a temporary advantage as prey ultimately depend on plants for food. So as plant species decline, so too would prey.

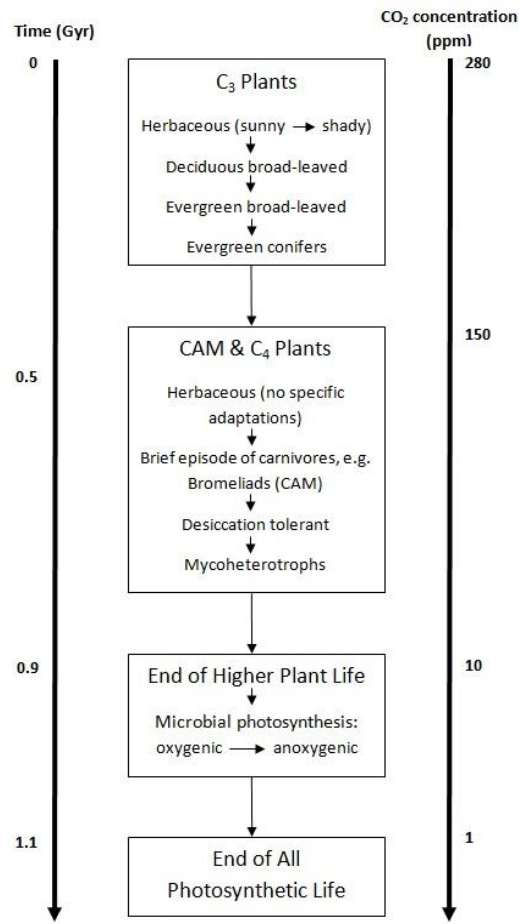

**Figure 4**: Plant extinction sequence. $C_3$ plants (plants that use a carbon fixation pathway that converts $CO_2$ into a 3-Carbon molecule) can survive until atmospheric $CO_2$ reaches 150 p.p.m. (Caldeira & Kasting 1992). These plants with lower $CO_2$ uptake rates at low $CO_2$ concentrations (from Mooney, 1972) die out first. The plants with $CO_2$ concentrating mechanisms, i.e. CAM (Crassulacean Acid Metabolism; those that use a carbon fixation pathway using external $CO_2$ in which the 4-carbon dicarboxylic malic acid is produced at night when stomata are open and stored until morning when stomata are closed. Malic acid is decarboxylated followed by assimilation of the resulting $CO_2$ by Rubisco) and $C_4$ (those that use a carbon fixation pathway that converts external $CO_2$ into a 4-carbon dicarboxylic molecule in the light, followed by movement of the 4-carbon acid into a compartment which has limited access to the atmosphere where the 4-carbon acid is decarboxylated and the resulting $CO_2$ assimilated by Rubisco) survive for longer (Caldeira & Kasting 1992; Winter & Smith 1996; Sage & Zhu 2011). Species decline until only those most tolerant to arid, low nutrient and high temperature conditions remain. In each case, larger plants would be more likely to face extinction as these have more mechanical and transport machinery to build and maintain, resulting in a greater loss of photosynthetically gained carbon via respiration. After atmospheric $CO_2$ falls below 10 p.p.m. only microbial photosynthesis remains viable until $CO_2$ levels fall too low for even this to take place.

Similar considerations apply to mycoheterotrophs, which are plants obtaining some or all of their organic carbon from symbiotic mycorrhizal fungi, which in turn obtain their organic carbon from soil organic carbon or from fully photosynthetic plants with which the fungi are also symbiotic (Leake & Cameron 2010). About 10% of vascular plants have a juvenile mycotrophic and subterranean phase growing from (usually) very small spores or seeds which depends on mycoheterotrophy (Leake & Cameron 2010). While almost all of these ultimately become photosynthetic autotrophs, about 450 species never develop photosynthesis and live as mycoheterotrophs through their life cycle (Leake & Cameron 2010). Such plants would only have a short time to live after the last autotrophic plants in a habitat died, using soil organic carbon that is no longer being replenished.

Adaptations that conserve water such as lower surface area per unit volume (as seen in cacti and other succulents, noting that this adaptation relates mainly to restricting water loss when photosynthesis using external $CO_2$ is not possible) or increased reflectivity via the use of epicuticular wax, such as in the case of *Dudleya brittonii*, may be more widely adopted as temperatures increase (Jones 2013; Sage & Zhu 2011). The drivers for these adaptations are not known, but the cause may have been a prolonged global dry spell which coincided with a global temperature drop and a fall in atmospheric $CO_2$ (Arakaki *et al.* 2011). Present-day $C_4$ plants are more rapid photosynthesisers in dry, high temperature, $CO_2$- and nitrogen-limited environments, suggesting adaptations like this may be employed as conditions for plants worsen.

The decline in plant species would lead to a lowering of atmospheric $O_2$ and $O_3$, which eventually fall to trace levels in only a few Myr once all plant life has disappeared (Walker 1991). Decaying plant life would produce methanethiol ($CH_3SH$) (see later). Animal and plant extinction sequences will inevitably overlap. Those animals that are in food chains dependent on live plants will begin declining shortly after the plants at the base of their food chains disappear. Larger animal species could survive for longer in the oceans where phytoplankton would outlast land plants. Animals that are not dependent on live plants could survive for longer. Termites, for example, are able to digest dead wood due to a symbiotic relationship with microorganisms in their guts (Geib *et al.* 2008), while some isopod crustaceans are able to digest dead wood without the aid of symbiotic microbes (King *et al.* 2010). Chemolithotrophically (chemosynthetically) symbiotic vetimentiferan worms (*Riftia*) found near hydrothermal vents may also be well suited to surviving longer than other animals, although these too would face extinction when the oxygen content of deep waters is depleted.

The decrease in atmospheric ozone would lead to less attenuation of DNA damaging ultraviolet radiation (UV), particularly UVC radiation (200-280 nm). The increased UV surface flux would be further enhanced by the decrease in atmospheric $CO_2$ levels, which, in the present-day atmosphere, scatter shorter wavelengths in the UVC range. This would lead to surface UV irradiation similar to that of the Archaean Earth, reaching $\sim 1 \times 10^{-1}$ $Wm^{-2}$ $nm^{-1}$ for longer UVC wavelengths (Cnossen *et al.* 2007) and a DNA-weighted dose up to a thousand times higher than on present-day Earth (Cockell & Raven 2007). If a similar irradiation environment to the Archaean is assumed, there are likely to still be environments where organisms could survive using DNA repair mechanisms or in shielded environments (Cockell & Raven 2007).

Land plants and oceanic phytoplankton produce significant amounts of the volatile compound isoprene (Seager *et al.* 2012). Isoprene has a very short atmospheric lifetime, being rapidly oxidised. However, in an increasingly anoxic atmosphere, isoprene may be a plausible atmospheric biosignature. Oceanic phytoplankton would be better able to survive for longer than terrestrial photosynthetic organisms due to their low $CO_2$ requirements and ability to extract carbon from bicarbonate (Raven *et al.* 2012). Therefore, if these organisms release isoprene at a rate similar to the present-day rate (0.31-1.09 Tg C $yr^{-1}$ - Gantt *et al.* 2009) this could allow a detectable level of isoprene to build up in the atmosphere, although the photodissociation of the increased atmospheric water vapour content could provide enough free oxygen to keep the lifetime of isoprene short (it has a lifetime of < 1 day in an oxygen-rich atmosphere - Palmer *et al.* 2003).

*b) Animals*
Decaying animals (as well as plants) release methanethiol ($CH_3SH$), a gas that can be spectrally inferred and has no known abiotic source. The extreme case of the extinction of all animals (vertebrates -> invertebrates) over a few Myr timescale could lead to a large release of $CH_3SH$ into the atmosphere. Assuming a biomass of plants and animals equivalent to present day values ($560 \times 10^9$ tonnes C - Groombridge & Jenkins 2002) and cell sulphur contents of 0.3-1% (Pilcher 2003), this allows for the potential release of $1.12 \times 10^{15}$ g of $CH_3SH$ (via conversion from methionine). If this were released at the current biological production rate ($3 \times 10^{12}$ g

yr$^{-1}$) it would all be released to the atmosphere within 350 years. CH$_3$SH itself would not be a useful biosignature as it has a very short atmospheric lifetime, being readily photodissociated. However, the dissociated methyl groups combine to form Ethane, which would have a much longer atmospheric lifetime under the increasingly anoxic conditions at the end of the age of animals, making it a potential biomarker under these circumstances (Domagal-Goldman *et al.* 2011). However, as with plant extinctions, the extinction of animal species is likely to occur over a much longer (10-100s Myr) timescale, which could lead to a much weaker biosignature associated with the event, e.g. a 100 Myr extinction time would lead to a smaller flux of ~10$^7$ g CH$_3$SH yr$^{-1}$.

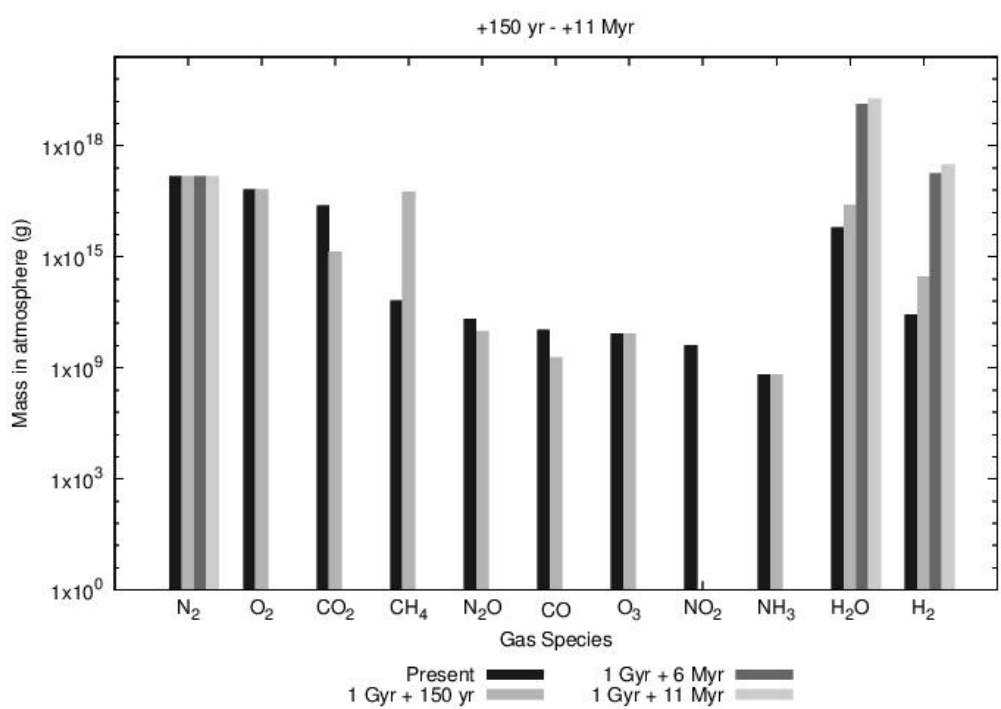

**Figure 5:** (A) Changes in gas species abundances in the atmosphere between 150 years and 11 Myr (+1.11 Gyr from present) after the extinction of higher plant species. The rapid disappearance of CO$_2$ and O$_2$ (and CO and O$_3$) is seen within 6 Myr as is the onset of runaway ocean evaporation and the associated increase in H$_2$ flux from increased photodissociation is clear. The large increase in CH$_4$ is associated with the decay of organic matter for the extreme case of rapid extinction. N$_2$O production drops off with declining O$_2$ levels as denitrification rates fall. This rapid decay halts the NH$_3$ flux, with the remaining atmospheric NH$_3$ being used up by ammonia oxidisers within 6 Myr. Under anoxic conditions, NO$_2$ production via the fixation of N$_2$ by lightning could be up to 2 orders of magnitude lower than production in the present day atmosphere. This would be more than balanced by the loss of NO$_2$ via photolysis, resulting in no net gain of NO$_2$.

*(ii) Chemolithotrophs in far-future refuge environments*

*a) High altitude*

Temperatures at high altitudes could remain low enough to permit liquid water pools for longer than at surface levels, providing refuges for life as surface temperatures increase. Additionally, any hot spot volcanoes (up-welling plumes from the deep mantle that occur in the middle of plates rather than at plate boundaries) that are present on the far-future Earth would provide refuges for simple microbial communities. Recent findings from a site at high altitude (>6000 m) in the Atacama Desert show that life in such environments uses the oxidation of carbon monoxide, ammonia or dimethyl sulphides to obtain energy (Lynch *et al.* 2012), producing various by-products such as $CO_2$, $N_2$, $SO_2$, methane sulphonate and sulphates (see [Table 1](#) for reactions). For oxidation reactions to be feasible, localised $O_2$ sources would be needed for such communities to exist in far-future refuge environments. A lesser energy gain per mol of substrate oxidised could also be obtained with other oxidants such as ferric iron and nitrate, but the availability of these tends to decrease with oxygen availability. As $O_2$ levels are reduced, iron will tend to be in a reduced state, making iron oxidation more favourable in a declining biosphere. Life is also known to exist around volcanic steam vents, for example, McMillan and Rushforth (1985) reported the presence of many species of diatom living in condensing steam around a steam vent near Kilauea Crater, Hawaii (as these are photosynthetic species, a sufficient $CO_2$ flux from the volcanic vent would be needed to support similar communities on the far-future Earth). Water vapour is the most abundant gas released via volcanic activity, which may mean that high altitude volcanic vents provide oases on the dry future Earth as steam condenses onto lower temperature surfaces.

There are over 1500 subaerial volcanoes that have been recently active, i.e. within the past 10,000 years (Siebert, Simkin & Kimberly 2010). Only approximately 49 volcanic sites are found at volcanic hotspots (Courtillot *et al.* 2003). Comparing hypsometric curves for Earth, Venus and Mars shows that, while there are clear differences in the surface area-elevation relationships for each planet, in all cases only < 1% of the surface area is found at elevation above 5 km.

*b) Caves*

Microbial life in cave environments on Earth today consists of chemolithotrophically based metabolisers and those that make use of organic carbon made available from photosynthesis, often being found in microbial mat communities. Life in far-future cave refuges is more likely to be chemolithotrophic due to the absence of photosynthetic carbon sources. An example of a cave ecosystem entirely based on chemosynthesis is found in Movile Cave, Romania. Here, life has been separated from the outside world for 5.5 Myr and has similarities to hydrothermal vent communities (Chen *et al.* 2009). The cave receives a high flow rate of hydrothermal water, which is rich in sulphides and methane; hence life in the cave is mainly based on sulphide and sulphur oxidation and methane oxidation. This produces by-products such as sulphuric acid, iron III (ferric) sulphate and carbon dioxide (see Table 1).

However, sulphur, sulphide and methane oxidation would only be possible to a limited extent on the largely anoxic far-future Earth. Sulphate and anaerobic methanotrophs have been found together in anoxic marine sediment environments (Orphan *et al.* 2001), suggesting a potential alternative for far-future cave ecosystems although this would not provide as much energy as oxidation reactions. Anammox bacteria, known to be high-temperature tolerant (Byrne *et al.* 2009), could potentially produce a hydrazine biosignature (assuming gas exchange with the surface) as hydrazine is a free intermediate in the anammox reaction (Strous *et al.* 2006).

*c) The deep subsurface*
The deep subsurface generally refers to depths of 50 m or more below the continental surface or ocean floor. The maximum inhabited depth is, at present, unknown - the deepest microbial life found to date was found at a depth of 5.3 km in igneous rock aquifers. Microbial subsurface life could use anaerobic methane oxidation, ferric iron reduction or sulphate reduction to metabolise (see Table 1). Hydrogen driven communities may also be present, although there is some debate as to whether enough abiotic hydrogen (produced by the radiolysis of water (Blair *et al.* 2007) or reactions between dissolved gases in magma, for example) could be produced to support them (Reith 2012). Originally it was assumed that subsurface life would be predominantly methanogenic; however, recent work suggests that this may not be the case due to the higher than expected counts of sulphate reducers found, suggesting a syntrophic biosphere of chemolithoautotrophs or organotrophs with fermenting bacteria (Teske 2005). Life in the deep subsurface would experience higher temperatures than surface dwelling life, as temperature increases with depth due to the geothermal temperature gradient of the Earth's crust. This means it is likely that this would be the first group of microorganisms on the dying Earth to disappear.

Detecting the presence of subsurface life may be more challenging than detecting life in other habitats. While life in the shallow subsurface could produce waste gases such as methane, which could percolate out into the atmosphere through rock or soil, deeper life would require the presence of deep faults in rocks for biomarker gases to escape into the atmosphere (Stevens 1997). However, in some cases it could be possible to detect subsurface life indirectly by looking at the surface features of the environments that support it (Hegde & Kaltenegger 2013). Temperature below the surface is determined by the surface temperature and increases with depth. The Earth's geothermal temperature gradient varies depending on rock type, but on average it increases by ~30° per km depth (Bohlen 1987). Although this mean

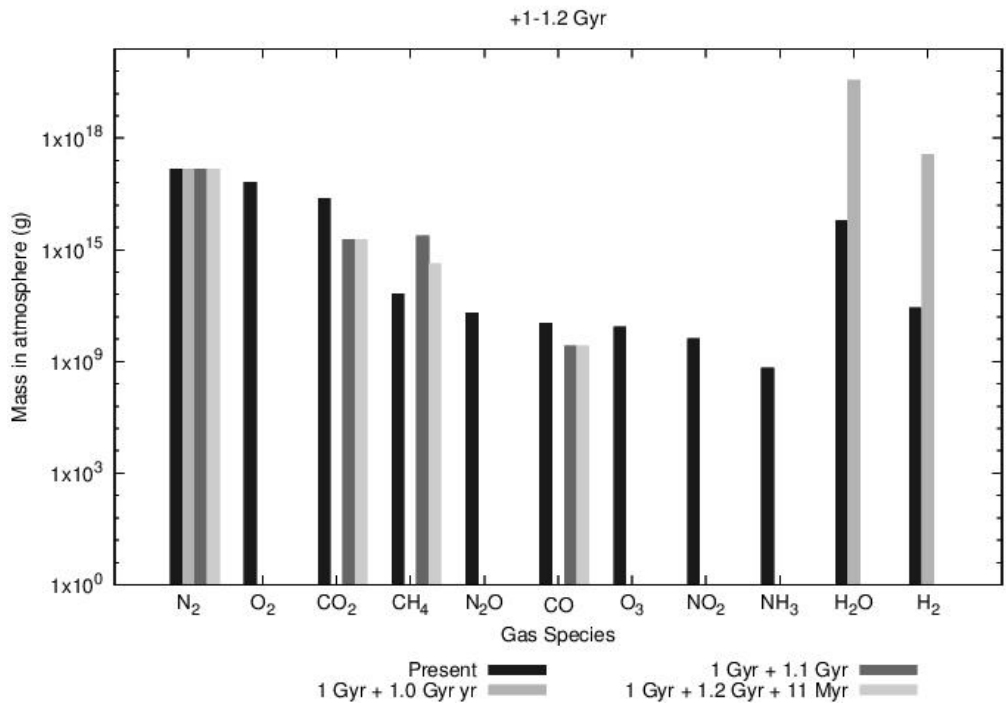

**Figure 6: (B)** Changes in gas species abundances in the atmosphere +2 Gyr from present. Atmospheric $H_2O$ levels fall as the oceans fully evaporate and the remaining water vapour photodissociates, with the liberated hydrogen escaping to space. The loss of atmospheric $H_2O$ slows silicate weathering rates, allowing $CO_2$ from volcanic sources to build up in the atmosphere once more. Methanogenic microorganisms make use of increased $CO_2$ availability, producing $CH_4$ as a by-product, which in turn is used by methanotrophs.

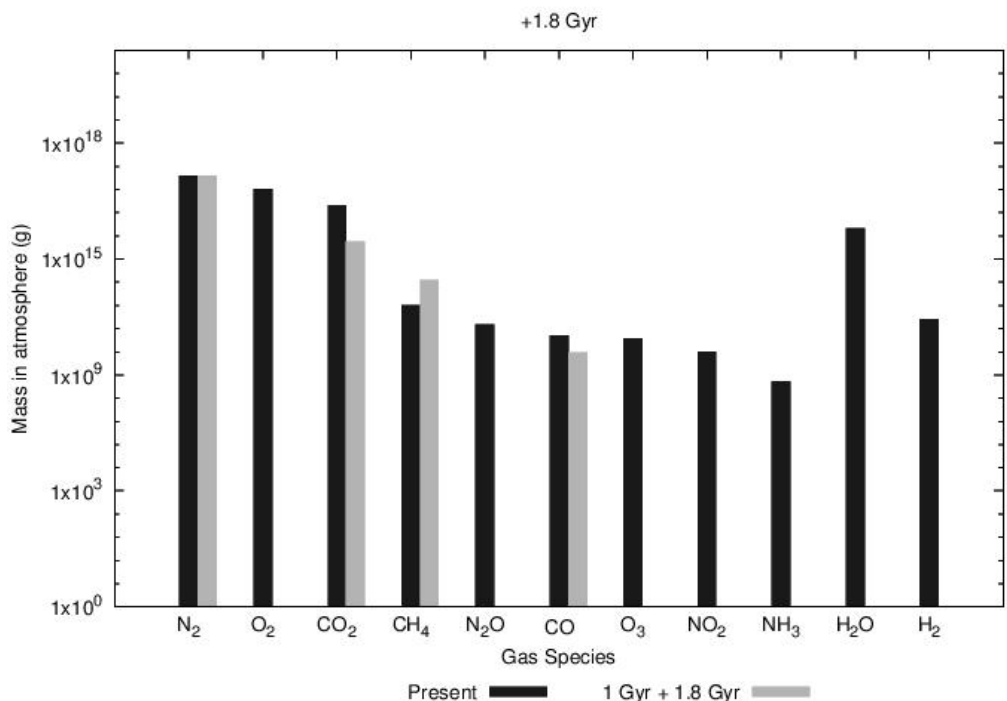

**Figure 7: (C)** Changes in gas species abundances in the atmosphere +2.8 Gyr from present. Only $CH_4$ produced by methanogenesis remains as a viable biosignature in a largely $N_2$:$CO_2$ atmosphere.

Page | 19

temperature gradient would preclude life once surface temperatures exceed the maximum temperature tolerance for life, the variable nature of the crust's temperature gradient could allow life to persist for longer than in any other refuge (O'Malley-James *et al.* 2013).

*(d) Remote detectability*

For atmospheric abundances similar to present-day levels, $CO_2$, $H_2O$ and $O_2$/$O_3$ would all be remotely detectable with low resolution spectrographs on an extrasolar Earth-like planet (Selsis 2004; Kaltenegger *et al.* 2010). The only case where atmospheric $O_2$ and $O_3$ would be detectable is in the beginning of the time period covered in (A) (see [Figure 5](#)). $H_2O$ would be detectable in cases (A) and (B) (Figures [5](#) and [6](#)). Estimated $CO_2$ levels are lower than current atmospheric levels, making the remote detection of this less likely in all three cases. Ethane related to the large-scale decay of plant and animal species, in the extreme case discussed in (A), is theoretically remotely detectable, but water vapour and other atmospheric aerosols could interfere with its detection (Domagal-Goldman *et al.* 2011). Although not as biologically relevant, $H_2$ should be detectable at the levels predicted in both cases (A) and (B), producing a particularly strong signal in case (B) (Miller-Ricci & Fortney 2010). $N_2$, despite its abundance, is spectrally inactive. It lacks any marked electronic transition (except at extreme UV wavelengths), but its presence can be inferred from collisions between $O_2$ and $N_2$ molecules when observing the transmission spectra of planetary atmospheres (Pallé *et al.* 2009). Oxygen collision complexes and molecular oxygen bands can then be used to derive an average atmospheric $N_2$ column density. In low $O_2$ atmospheres, detecting $N_2$ may therefore be more challenging. From the minimum abundance limits for detection described in Selsis (2004), only $CH_4$ would be potentially detectable in cases (A), (B) and (C) - Figures [5](#), [6](#) and [7](#) - with $NH_3$ being detectable in case (A). All other atmospheric species fall below the minimum abundance thresholds in each case. At current $CH_4$ levels, the main $CH_4$ spectral feature is found in the mid-IR wavelength region (at ~7 μm), but this is largely obscured by adjacent $H_2O$ and $N_2O$ features (Des Marais *et al.* 2002). The predicted far-future $CH_4$ levels are approximately an order of magnitude greater than present day levels, which, combined with negligible atmospheric $H_2O$ and $N_2O$ levels would make the 7 μm feature clearer and more readily remotely detectable. Upcoming and planned future missions such as the James Webb Space Telescope (Gardner *et al.* 2006), Terrestrial Planet Finder (Lawson *et al.* 2008) and the Exoplanet Characterisation Observatory (EChO) (Tinetti *et al.* 2012) would all be looking in appropriate wavelength ranges to detect such a feature (although EChO would be limited to super-Earth ( > 10 Earth-mass) planets), with follow-up observations being possible with future ground-based instruments such as the European Extremely Large Telescope (Gilmozzi & Spyromilio 2007). However, a positive $CH_4$ detection may not necessarily indicate the presence of life. $CH_4$ can be released to the atmosphere abiotically via volcanic activity and serpentinization, for example (Oze *et* *al.* 2012; Burton *et al.* 2013). Building a case for a biological $CH_4$ source on a late-habitable-stage Earth-like planet may be easier than for younger habitable planets due to the slowing of carbon recycling as a result of the slowing of plate tectonics and a lower abundance of water to drive serpentinization reactions. Recent experimental work also suggests that a low atmospheric $H_2/CH_4$ ratio would support the case that $CH_4$ is being produced life rather than via serpentinization (Oze *et al.* 2012). The potentially detectable biosignature gases are summarised in [Table 4](Table 4).

| Stage of biosphere decline | Biosignature gases |
|---|---|
| A (1 - 1.11 Gyr) ([Figure 5](Figure 5)) | $O_2$, $O_3$, $H_2O$, $C_2H_6$, $NH_3$, $CH_4$ |
| B (2 - 2.2 Gyr) ([Figure 6](Figure 6)) | $H_2O$, $CH_4$ |
| C (2.8 Gyr) ([Figure 7](Figure 7)) | $CH_4$ |

**Table 4:** Summary of potentially detectable biosignature gases.

*(iii) Other biosignatures*

*a) Clouds*
A potential refuge for microbial life in the far future would be in the atmosphere where temperatures would be cooler than at maximum land elevation. It has been postulated that if life existed on a more clement early Venus, it could have found its way into the venusian atmosphere and could be making a living in the upper cloud decks where temperatures fall within a habitable range (Schulze-Makuch 2004).

If a similar scenario were to occur when Earth's oceans evaporate, one of the final refuges for life may be in the air. Microbial life is known to exist in the atmosphere today, although it is not yet known whether they are just in transit or whether they are actively metabolising and reproducing in the atmosphere. Airborne microorganisms can influence cloud formation (see later discussion), suggesting that if biological cloud formation can be distinguished from abiotic cloud formation, clouds could be used as bio-indicators of an aerial biosphere.

Biological particles have little to no effect on cirrus clouds (thin, wispy clouds above altitudes of 5 km) (Möhler *et al.* 2007). Different types of cloud can be distinguished using information on cloud-top pressure (the atmospheric pressure at the cloud top) and optical thickness. Cirrus clouds (the most common type of cloud on Earth, making up 25% of the total global cloud cover and playing a large role in Earth's radiation budget) have high cloud-top pressures and low optical thickness, while convective clouds (responsible for most precipitation in the tropics) have a high cloud-top pressure and high optical thickness (Durieux *et al.* 2003). It is feasible that these properties could be detected in the atmospheric spectra of exoplanets, although if the cloud-top is at a deep layer in the atmosphere and the optical thickness of the cloud layer is high, direct detection is not possible, but a lower limit on cloud-top pressure can be determined (Benneke & Seager 2012).

In a recent study of the affect of airborne microbial populations on cloud formation, DeLeon-Rodriguez *et al.* (2013) found that 20% of particles sampled in the upper troposphere were bacterial (the remainder being salt, dust etc.). If we assume that generally, 20% of cloud seeding particles are microbial this could make cloud cover itself a biosignature on inhabited, arid planets if the expected abiotic cloud coverage could be estimated. For largely arid, desert planets, high desert dust levels in the atmosphere could influence the predominant cloud type. Saharan desert dust, for example, acts as small cloud condensation nuclei, causing clouds composed of small water droplets (< 14 μm radii) to form, reducing coalescence and suppressing precipitation (Rosenfeld *et al.* 2001). Cloud water droplet size could be determined via remote sensing of exoplanet atmospheres using the rainbow scattering effect (Bailey 2007); hence, unexpectedly large cloud droplets in the atmospheres of arid planets could provide indirect evidence of microbial cloud seeding.

*b) Leaf area index*

A decrease in vegetation during the stage in which plant extinction occurs could produce detectable spectral signatures. Remote sensing can be used to estimate vegetation coverage from the leaf area index (LAI), defined as green leaf area/surface area. Spectral reflectance in the near infrared and visible wavelength regions is used to determine the normalised difference vegetation index (NDVI)

$$NDVI = \frac{\rho_{NIR} - \rho_{vis}}{\rho_{NIR} + \rho_{vis}} \qquad (1)$$

from which an estimate of LAI (locally or globally) can be obtained (see, for example Deng *et al.* (2006), Ganguly *et al.* (2012)). As plant extinction rates increase, the LAI should decrease as leaf area decreases, potentially leading to a less intense red edge spectral feature. Filella & Peñuelas (1994) found that the area of red edge peak changed with LAI as LAI determines the ratio of near-IR and red reflectance. $C_4$ plants have been shown to exhibit greater leaf area than $C_3$ plants grown under the same conditions (Anten *et al.* 1995), suggesting the presence of more rapid photosynthesisers could delay a reduction in the spectral red edge signal. However, as plant species become extinct, vegetation will become increasingly sparse, with LAI values approaching high desert values (< 1 - Running *et al.* 1986) and the LAI - NDVI relation may no longer hold. When vegetation cover becomes too low, the observed spectrum is dominated by soil and rock which have varied red-near-IR slopes, which alter the measured vegetation indices (Huete *et al.* 1985; Elvidge & Lyon 1985). Additionally, as vegetation becomes more adapted to arid conditions it is likely to incorporate adaptations that decrease the amount of visible light absorbed, making it harder to detect.

Increasingly sparse vegetation will lessen the intensity of the red edge signature, which will eventually disappear completely with the end of all plant life. Warming temperatures lead to a poleward expansion of the boundaries of equatorial Hadley cells (currently at ±30° latitude), which has been linked to the simultaneous poleward expansion of subtropical dry zones (Lu *et al.* 2007). The meridional extent of the equatorial Hadley cells can be approximated using

$$\phi_H \cong \left(\frac{5gH_t\Delta h}{3\Omega^2 a^2 \theta_0}\right)^{\frac{1}{2}}$$

where g is the acceleration due to gravity, $H_t$ is the height of the tropical tropopause, $\Delta h$ is the equator-to-pole difference in radiative-equilibrium potential temperature, $\Omega$ is the rate of rotation of the Earth, a is the radius of the Earth and $\theta_0$ is the global mean temperature (Showman *et al.* 2010). Using the model results for temperature evolution and accounting for the slowing of the Earth's rotation due to tidal interactions with the Moon (increasing the day length by $1.70\times10^{-5}$ s yr$^{-1}$ - Stephenson 1997), $\phi_H$ can be used to estimate the expansion of the subtropical dry zones, giving an approximate timescale for the loss of abundant vegetation. Within 0.3 Gyr equatorial Hadley cells would have expanded to ~40° north and south of the equator, increasing the arid fraction of the surface by 25%, doubling the current total area of arid land. By assuming plants in arid regions are too sparsely spread to contribute to the red edge signature, this alone would reduce the strength of the signature. However, Earth is currently half-way through a supercontinent cycle, with another supercontinent expected to be formed within the next 250 Myr (Yoshida & Santosh 2011). The location of this landmass would greatly influence environmental conditions. An equatorial supercontinent would have a more stable climate, but would likely have large dry, arid interiors, whereas polar supercontinents would experience extreme temperature fluctuations, except in coastal regions (Williams & Kasting 1997). Either case would lead to large areas of land with only sparse vegetation coverage, effectively ending the distinctive vegetation red edge signature. Supercontinent formation and break-up have been linked to past mass extinction events (Santosh 2010). Increased competition between species due to the merging of continental landmasses and the possibility of increased volcanic emissions from deep mantle superplumes, combined with generally less hospitable conditions on the warming Earth could lead to a mass extinction event for plants and animals from which they may not be able to recover.

*iv) Biosignatures of dying biospheres on non-Earth-like planets*

Habitable planets that are not direct Earth analogues would probably exhibit different biosignatures associated with dying planetary biospheres when compared to the case of Earth analogue planets nearing the end of their habitable lifetimes.

Smaller, colder Mars-like planets would end their habitable lifetimes in a cold, arid state, with the best habitats for any remaining life being in the warmer subsurface regions. Outgassing from the near subsurface could be one of the indicators of ongoing biological activity on such planets.

For example, there is a claim that methane has been detected in plumes emanating from three distinct areas on Mars (Mumma *et al.* 2003). As the lifetime of methane in Mars' atmosphere is relatively short (Elwood Madden *et al.* 2011), the detection of methane implies continual replenishment of the gas. Methane in Earth's atmosphere is largely biological in origin; hence, it has been suggested that this may be an indicator of ongoing biological processes on the planet (Mumma *et al.* 2003). However, there are also abiotic mechanisms, such as the release of ancient methane trapped beneath the surface, that could be responsible for the atmospheric methane (McMahon *et al.* 2013) and the actual detection of methane on Mars has been called into question (Zahnle *et al.* 2011; Webster *et al.* 2013).

Similarly, hotter, inner-habitable zone Venus-like planets would end their habitable lifetimes in a hotter and potentially geologically active state (the *Venus Express* spacecraft found evidence of recent volcanism on Venus - Smrekar *et al.* 2010). The habitable lifetimes of such planets would likely be much shorter than Earth-like planets (if warm oceans were present on early Venus, its habitable lifetime would correspond to the lifetime of liquid surface water on the planet - estimates for this range from 600 Myr (Kasting 1988) to a few Gyr (Grinspoon & Bullock 2003)). The upper atmosphere would likely be the best final refuge for any life that developed on such planets. For example, the possibility of microbial life in the cloud decks of Venus has long been postulated (Sagan & Morowitz 1967; Sagan & Saltpeter 1976; Cockell 1999). It may be that any microbial life that developed on the early Venus was able to migrate upwards as the planet's oceans evaporated, perhaps using sulphur allotropes in the atmosphere to survive (Schulze-Makuch *et al.* 2004), with the presence of such life potentially being inferred from unexpected UV radiation absorption (a UV-transparent biosphere would be less favourable than one in which UV screening and repair mechanisms are developed - Cockell & Airo 2002) and atmospheric chemical disequilibrium.

**CONCLUSIONS**

The term "Earth-like", when applied to atmospheric spectra or life tends to imply conditions similar to those on the present-day Earth. Earth's atmosphere and life inhabiting the planet have evolved considerably over time and will continue to do so into the future, which means that for a lot of Earth's habitable lifetime conditions would not be described as Earth-like when compared to the present environment. The implications for the search for extraterrestrial life are that the search criteria may need to be broadened to improve the chances of successfully detecting life.

The extinction of plants and animals caused by a warming planet as solar luminosity increases and the subsequent evaporation of the oceans will lead to different atmospheric composition and biosignatures when compared to the present. In the dry, microbial world left behind as Earth loses its water, observed spectra would once again change, as the spectra of a desert world with little or no ocean cover would exhibit light curves with much greater reflectivity than ocean and vegetation covered planets (Ford *et al.* 2003).

# APPENDIX

*Equations for silicate weathering and carbonate deposition*:

Weathering on land:
$CaSiO_3 + 2\ CO_2 + 3\ H_2O \rightarrow Ca^{2+} + 2HCO_3^- + Si(OH)_4$
After movement down rivers, in ocean:
$Ca^{2+} + 2HCO_3^- \rightarrow CaCO_3\downarrow + CO_2 + H_2O$
$Si(OH)_4 \rightarrow SiO_2\downarrow + 2H_2O$

After subduction (and after much delay) volcanism:
$CaCO_3 + SiO_2 \rightarrow CaSiO_3 + CO_2$

If the bicarbonate stayed in solution in the ocean more $CO_2$ would be drawn down in weathering, but it would eventually precipitate (even in the absence of calcifying organisms) as $CO_3$ with production of half of the $CO_2$ take up in the initial weathering.

# ACKNOWLEDEMENTS

The authors wish to thank Stefanie Lutz and Dirk Schulze-Makuch for helpful discussions and an anonymous reviewer for comments that improved the manuscript. The University of Dundee is a registered Scottish charity, No. SC015096. JTO acknowledges an STFC Aurora grant.